\documentclass[preprint,aps,superscriptaddress,nofootinbib,tightenlines]{revtex4}
\usepackage{epsfig}
\usepackage{bm}

\usepackage{epsfig}
\newcommand{\beq}{\begin{equation}}
\newcommand{\eeq}{\end{equation}}
\newcommand{\bea}{\begin{eqnarray}}
\newcommand{\eea}{\end{eqnarray}}

\def\OMIT#1{{}}

\def\si{^1 \hskip -0.03in S _0}
\def\siii{^3 \hskip -0.025in S _1}
\def\diii{^3 \hskip -0.03in D _1}

\begin{document}
\begin{figure}[!t]
\vskip -1.5cm
\leftline{
{\epsfxsize=1.8in \epsfbox{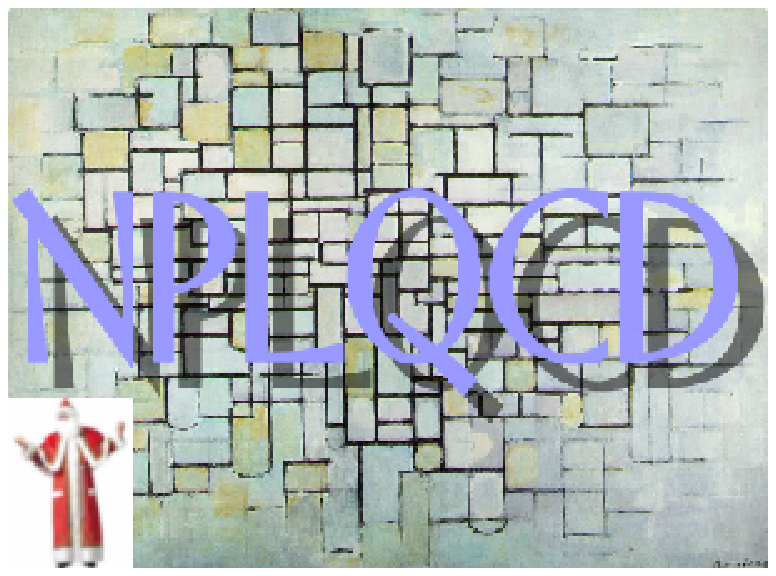}}}
\vskip 1.5cm
\end{figure}

\preprint{\vbox{
\hbox{UNH-06-07}
\hbox{UCRL-JRNL-226919}
\hbox{UMD-40762-364}
\hbox{JLAB-THY-06-600}
\hbox{NT@UW-06-031}
\hbox{UG-06-10}
}}

\vskip 1.5cm

\title{Hyperon-Nucleon Scattering from Fully-Dynamical Lattice QCD}
\author{Silas R.~Beane}
\affiliation{Department of Physics, University of New Hampshire,
Durham, NH 03824-3568.}
\author{Paulo F.~Bedaque}
\affiliation{Department of Physics, University of Maryland, College Park, MD 20742-4111.}
\author{Thomas C.~Luu}
\affiliation{N Division, Lawrence Livermore National Laboratory, Livermore, CA 94551.}
\author{Kostas Orginos}
\affiliation{Department of Physics, College of William and Mary, Williamsburg,
  VA 23187-8795.}
\affiliation{Jefferson Laboratory, 12000 Jefferson Avenue, 
Newport News, VA 23606.}
\author{Elisabetta Pallante}
\affiliation{Institute for Theoretical Physics, University of Groningen,
Nijenborgh 4,\\ 9747 AG  Groningen, The Netherlands.}
\author{Assumpta Parre\~no}
\affiliation{Departament d'Estructura i Constituents de la Mat\`{e}ria,
Universitat de Barcelona, \\ E--08028 Barcelona, Spain.}
\author{Martin J.~Savage}
\affiliation{Department of Physics, University of Washington, 
Seattle, WA 98195-1560.\\
\qquad}
\collaboration{ NPLQCD Collaboration }
\noaffiliation
\vphantom{}
\vskip 0.8cm
\begin{abstract}
We present results of the first fully-dynamical lattice QCD
determination of hyperon-nucleon scattering.  One s-wave phase shift
was determined for $n\Lambda$ scattering in both spin-channels at pion
masses of $350$, $490$, and $590~{\rm MeV}$, and for $n\Sigma^-$
scattering in both spin channels at pion masses of $490$, and
$590~{\rm MeV}$.  The calculations were performed with domain-wall
valence quarks on dynamical, staggered gauge configurations with a
lattice spacing of $b\sim0.125~{\rm fm}$.
\end{abstract}
\pacs{}
\maketitle


\subsection{Introduction}

\noindent In high-density baryonic systems, the large value of the
Fermi energy may make it energetically advantageous for some of the
nucleons to become hyperons, with the increase in rest mass being more
than compensated for by the decrease in Fermi energy. This is
speculated to occur in the interior of neutron stars, but a
quantitative understanding of this phenomenon depends on knowledge of
the interactions among the hadrons in the medium.  In contrast to
nucleon-nucleon (NN) interactions, where the wealth of experimental
data has allowed for the construction of high-precision potentials,
the hyperon-nucleon (YN) interactions are only very-approximately
known.  Experimental information about the YN interaction comes mainly
from the study of hypernuclei~\cite{hypernuclei-review,Hashimoto:2006aw}, the analysis
of associated $\Lambda$-kaon and $\Sigma$-kaon production in NN
collisions near threshold~\cite{Ba98,Bi98,Se99,Ko04,AB04,GHHS04}, and
hadronic atoms~\cite{Batty:1997zp}.  There are a total of 35
cross-sections measurements~\cite{NNonline} of the processes $\Lambda
p\rightarrow \Lambda p$, $\Sigma^-p\rightarrow\Lambda n$, $\Sigma^+
p\rightarrow\Sigma^+ p$, $\Sigma^- p\rightarrow\Sigma^- p$ and
$\Sigma^- p\rightarrow\Sigma^0 n$, and unsurprisingly, the extracted
scattering parameters are highly model dependent.  The theoretical
study of YN interactions is hindered by the lack of experimental
guidance.  The ``realistic'' potentials developed by the
Nijmegen~\cite{nij99,nij06} and J\"ulich~\cite{HHS89,RHKS96,HM05} groups are
just two examples of phenomenological models based on meson exchange.
These are soft-core potentials with one-boson exchange models of the
NN interaction.  Since $SU(3)$ flavor symmetry is broken by the
differences in the quark masses, the corresponding couplings are not
completely determined by the NN interaction and are instead obtained
by a fit to the available data.  In Ref.~\cite{nij99,nij06}, for example,
six different models are constructed, each describing the available YN
cross-section data equally well, but predicting different values for
the phase shifts.  The effective field theory
approach~\cite{savage-wise,KDT01,Hammer02,BBPS05,PHM06} is less
developed and suffers from a large number of couplings that need to be
fit to the data.

In view of the large uncertainties in the YN scattering amplitudes and
their importance for modeling neutron stars and the study of
hypernuclei, a first-principles QCD calculation of YN scattering is
highly desirable.  The only way to achieve this is through numerical
calculations using lattice QCD.  In a previous paper~\cite{BBPS05},
some of the present authors outlined a program to address this issue
with a combination of lattice calculations and the use of effective
field theories.  This paper reports on the first results of the
implementation of this program. In particular, we compute low-energy
s-wave phase shifts for YN scattering in the $\si$ channel and
$\siii-\diii$ coupled-channels at particular energies, using
L\"uscher's finite-volume
method~\cite{Hamber:1983vu,Luscher:1990ux,Beane:2003da}.  This is a
straightforward extension of work by some of the present authors on NN
scattering from lattice QCD~\cite{Beane:2006mx}.  The calculations
were performed in the isospin limit on the coarse MILC
lattices~\cite{Bernard:2001av} for pions with masses of $\sim 290$
MeV, $\sim 350$ MeV, $\sim 490~{\rm MeV}$ and $\sim 590~{\rm MeV}$.
However, we do not attempt to extrapolate to the physical pion mass as
it is likely that all but one of the data points lies outside the
regime of applicability of the YN EFT's.

\subsection{Methodology and Details of the Calculation}

\noindent Lattice QCD calculations of the {\it interactions} among
hadrons are notoriously difficult and require circumventing the
Maiani-Testa theorem~\cite{Maiani:1990ca}, which states that one
cannot compute Green's functions at infinite volume on the lattice and
recover S-matrix elements except at kinematic thresholds.
The s-wave scattering amplitude for two particles below
inelastic thresholds can be determined using L\"uscher's
method~\cite{Hamber:1983vu,Luscher:1990ux,Beane:2003da}, which entails a measurement of one or
more energy levels of the two-particle system in a finite volume. 
Our computation uses the mixed-action lattice QCD scheme developed by
LHPC~\cite{Renner:2004ck,Edwards:2005kw} which places domain-wall
valence quarks from a smeared-source on $N_f=2+1$
asqtad-improved~\cite{Orginos:1999cr,Orginos:1998ue} MILC
configurations generated with rooted staggered sea
quarks~\cite{Bernard:2001av} that are hypercubic-smeared
(HYP-smeared)~\cite{Hasenfratz:2001hp,DeGrand:2002vu,DeGrand:2003in,Durr:2004as}.
In the generation of the MILC configurations, the strange-quark mass
was fixed near its physical value, $b m_s = 0.050$, (where
$b=0.125~{\rm fm}$ is the lattice spacing) determined by the mass of
hadrons containing strange quarks.  The two light quarks in the
configurations are degenerate (isospin-symmetric), with masses
$bm_l=0.007$, $0.010$, $0.020$ and $0.030$. As was shown by
LHPC~\cite{Renner:2004ck,Edwards:2005kw}, HYP-smearing allows for a
significant reduction in the residual chiral symmetry breaking at a
moderate extent $L_s = 16$ of the extra dimension and domain-wall
height $M_5=1.7$.  Using Dirichlet boundary conditions we reduced the
original time extent of 64 down to 32. This allowed us to recycle
propagators computed for the nucleon structure function calculations
performed by LHPC. For bare domain-wall fermion masses we used the
tuned values that match the staggered Goldstone pion to few-percent
precision~\cite{Renner:2004ck,Edwards:2005kw}.  The parameters used in
the propagator calculation can be found in Ref.~\cite{Beane:2006gj}. 
All propagator calculations were performed using the Chroma software
suite~\cite{Edwards:2004sx,McClendon:2001aa}.  Eight propagators per
configuration were computed at distinct source points on the lattice.

We found that the cleanest method for extracting the energy-difference
between the YN state, and the mass of an isolated nucleon and an
isolated hyperon, was by forming the ratio of correlation functions
\begin{eqnarray}
G^{S}_{YN}(t) & = & C_{YN}^{S}(t)/\left( C_{Y}(t) C_{N}(t)  \right)
\rightarrow\
e^{-\Delta E_{YN} t}
\ \ ,
\label{eq:ratio}
\end{eqnarray}
where $S$ denotes spin.  At large times, this ratio depends
exponentially upon the ground-state energy shift of the YN system due to 
interactions.  The single nucleon correlator is
\begin{eqnarray}
C_{N}(t) & = & \sum_{\bf x}
\langle N(t,{\bf x})\ N^\dagger(0, {\bf 0})
\rangle
\ \ \ ,
\label{N_correlator} 
\end{eqnarray}
and the single hyperon correlator has an analogous form.  The YN
correlator that projects onto the s-wave state in the continuum limit
is
\begin{eqnarray}
& & C_{YN}^{S}(t) \ =\  
X_{\alpha\beta\sigma\rho}^{ijkl}
\nonumber\\
& & \qquad
\sum_{\bf x , y}
\langle Y^\alpha_i(t,{\bf x})N^\beta_j(t, {\bf y})
Y^{\sigma\dagger}_k (0, {\bf 0})N^{\rho\dagger}_l(0, {\bf 0})
\rangle
\ \ \ , 
\label{YN_correlator} 
\end{eqnarray}
where $\alpha,\beta,\sigma,\rho$ are isospin-indices and $i,j,k,l$ are
Dirac-indices.  The tensor $X_{\alpha\beta\sigma\rho}^{ijkl}$ has
elements that produce the correct spin-isospin quantum numbers for a
hyperon and nucleon in an s-wave.  The summation over ${\bf x}$ (and
${\bf y}$) corresponds to summing over all the spatial lattice sites,
thereby projecting onto the momentum ${\bf p}={\bf 0}$ state of each
particle seperately.  The interpolating field for the proton is
$p_i(t,{\bf x}) = \epsilon_{abc} u^a_i(t, {\bf x}) \left( u^{b T}(t,
{\bf x}) C\gamma_5 d^c(t, {\bf x})\right)$, and similarly for the
neutron and hyperons.  We have used an interpolating field
$n\times\Sigma^-$ to determine the energy-eigenvalues of the s-wave
strangeness $=1$, isospin $={3\over 2}$ eigenstates in both spin
channels, and an interpolating field $n\times\Lambda$ to determine the
energy-eigenvalues of the s-wave strangeness $=1$, isospin $={1\over
2}$ eigenstates in both spin channels.

Once the energy shift due to the YN interactions has been computed,
the real part of the inverse scattering amplitude is determined via
the L\"uscher formula~\cite{Hamber:1983vu,Luscher:1990ux,Beane:2003da}.  To extract
$p\cot\delta(p)$, where $\delta(p)$ is the phase shift, the magnitude of
the center-of-mass momentum, $p$, is extracted from the energy shift,
$\Delta E_{YN}=\sqrt{p^2+M_Y^2}+\sqrt{p^2+M_N^2}-M_Y-M_N$, and
inserted into:
\begin{eqnarray}
p\cot\delta(p) \ =\ {1\over \pi L}\ {\bf
  S}\left(\,\frac{p L}{2\pi}\,\right)
\ \ ,
\label{eq:energies}
\end{eqnarray}
which is valid below the inelastic threshold. The regulated three-dimensional sum is
\begin{eqnarray}
{\bf S}\left(\,{\eta}\, \right)\ \equiv \ \sum_{\bf j}^{ |{\bf j}|<\Lambda}
{1\over |{\bf j}|^2-{\eta}^2}\ -\  {4 \pi \Lambda}
\ \ \  ,
\label{eq:Sdefined}
\end{eqnarray}
where the summation is over all triplets of integers ${\bf j}$ such that $|{\bf j}| < \Lambda$ and the
limit $\Lambda\rightarrow\infty$ is implicit.

\subsection{Results}

\begin{figure}[!ht]
\vspace{0.2in}
\includegraphics[width = .32\textwidth,angle=0] {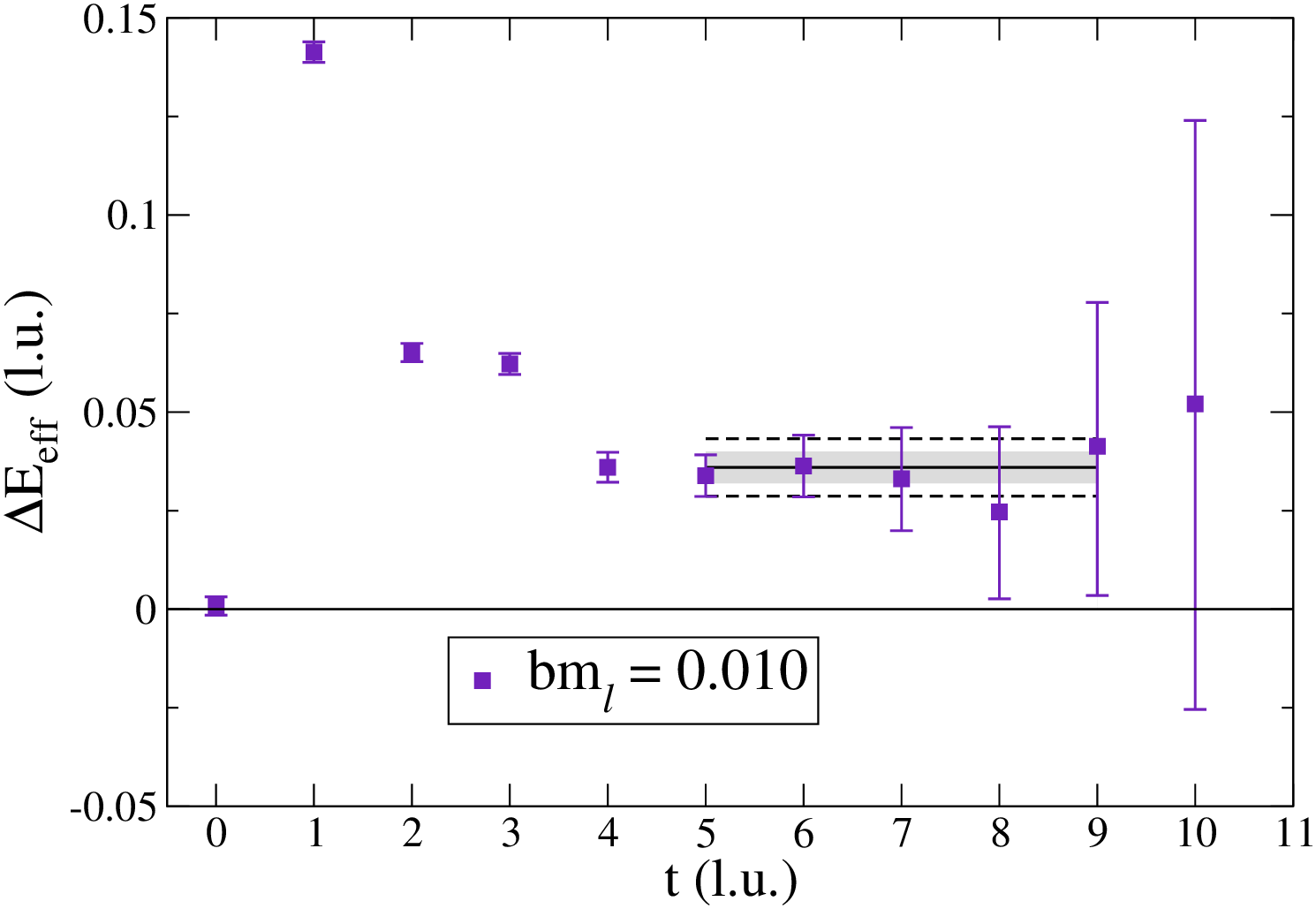} \hfill
\includegraphics[width = .32\textwidth,angle=0] {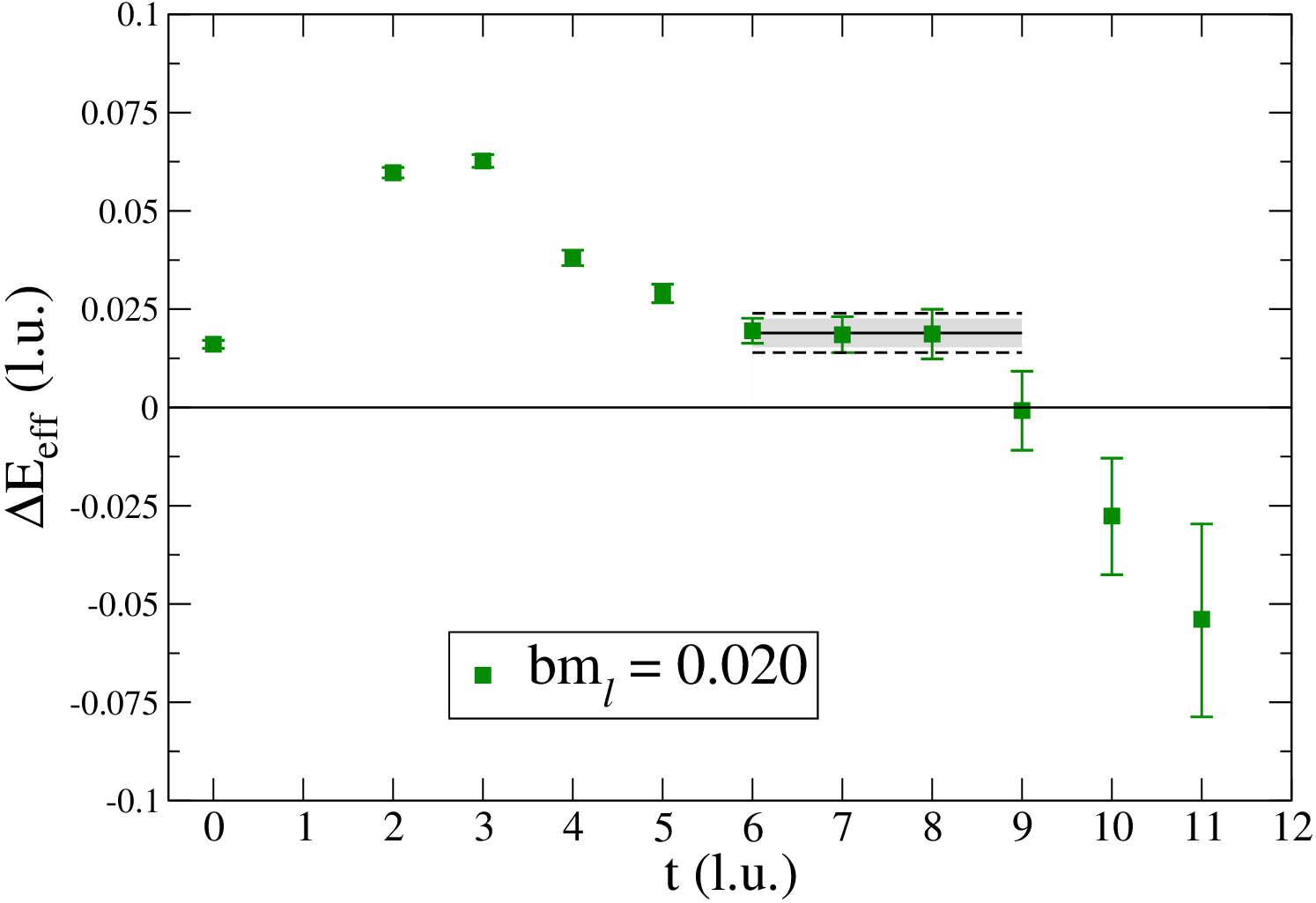} \hfill
\includegraphics[width = .32\textwidth,angle=0] {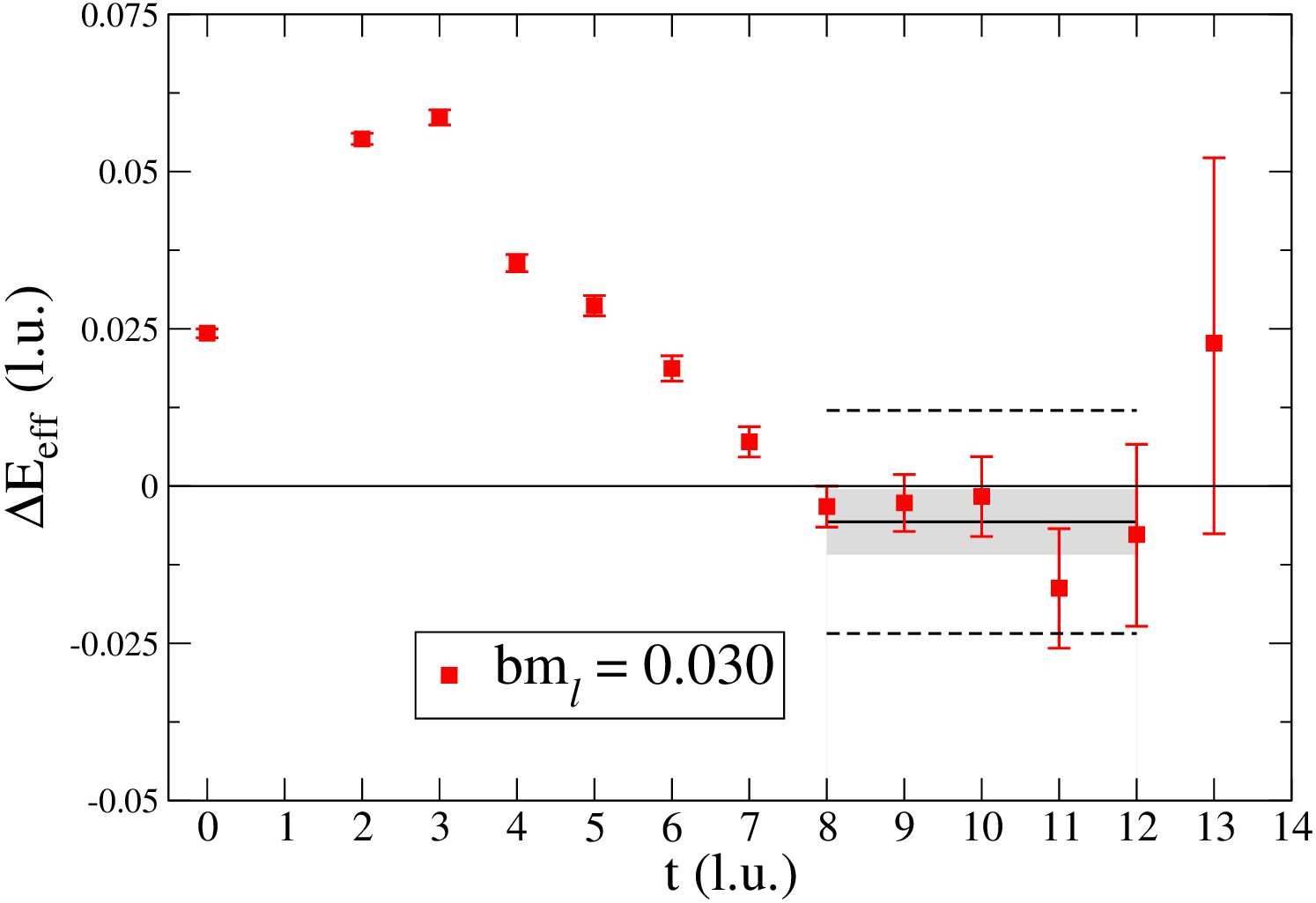}
\caption{Effective mass plots for  $n\Lambda$ in the $\si$-channel
at pion masses of $m_\pi\sim 350~{\rm MeV}$ (left panel),
$m_\pi\sim 490~{\rm MeV}$ (center panel)
and $m_\pi\sim 590~{\rm MeV}$ (right panel).
The straight line and shaded region correspond to the extracted energy shift
and associated uncertainty. The dashed lines correspond to the statistical
and systematic errors added linearly.}
\label{fig:effmassNL1s0} 
\end{figure}

\begin{figure}[!ht]
\includegraphics[width = .32\textwidth,angle=0] {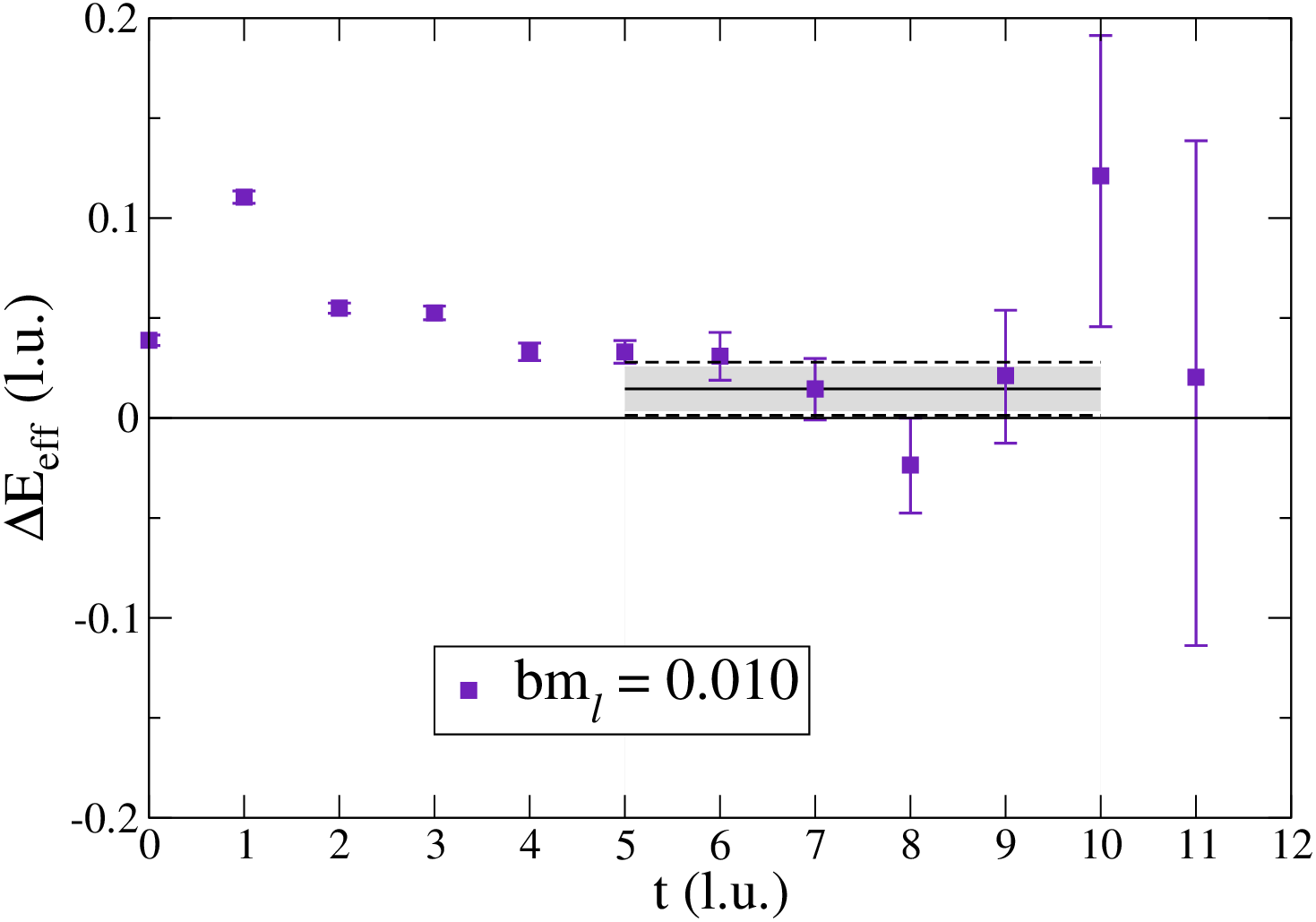} \hfill
\includegraphics[width = .32\textwidth,angle=0] {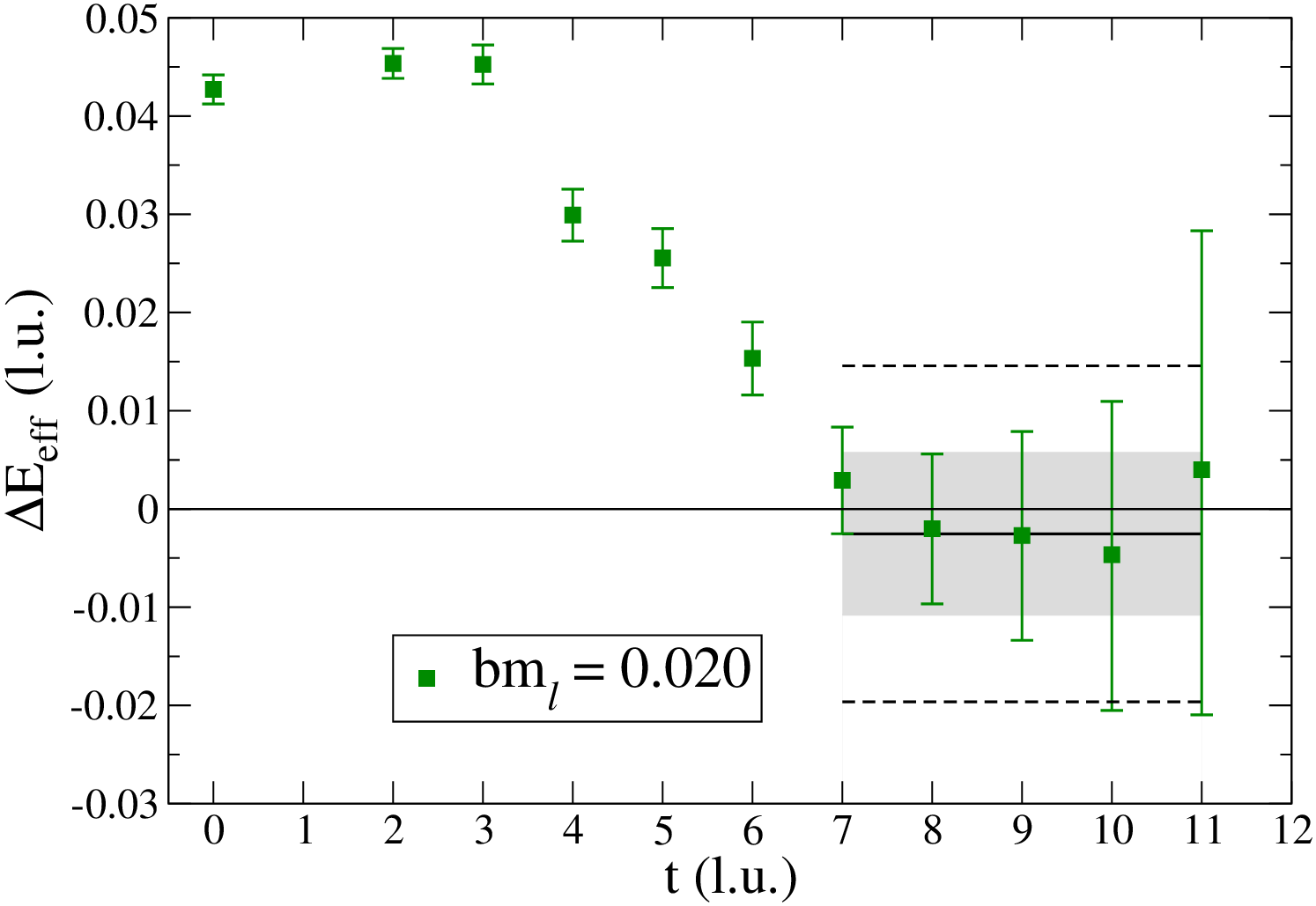} \hfill
\includegraphics[width = .32\textwidth,angle=0] {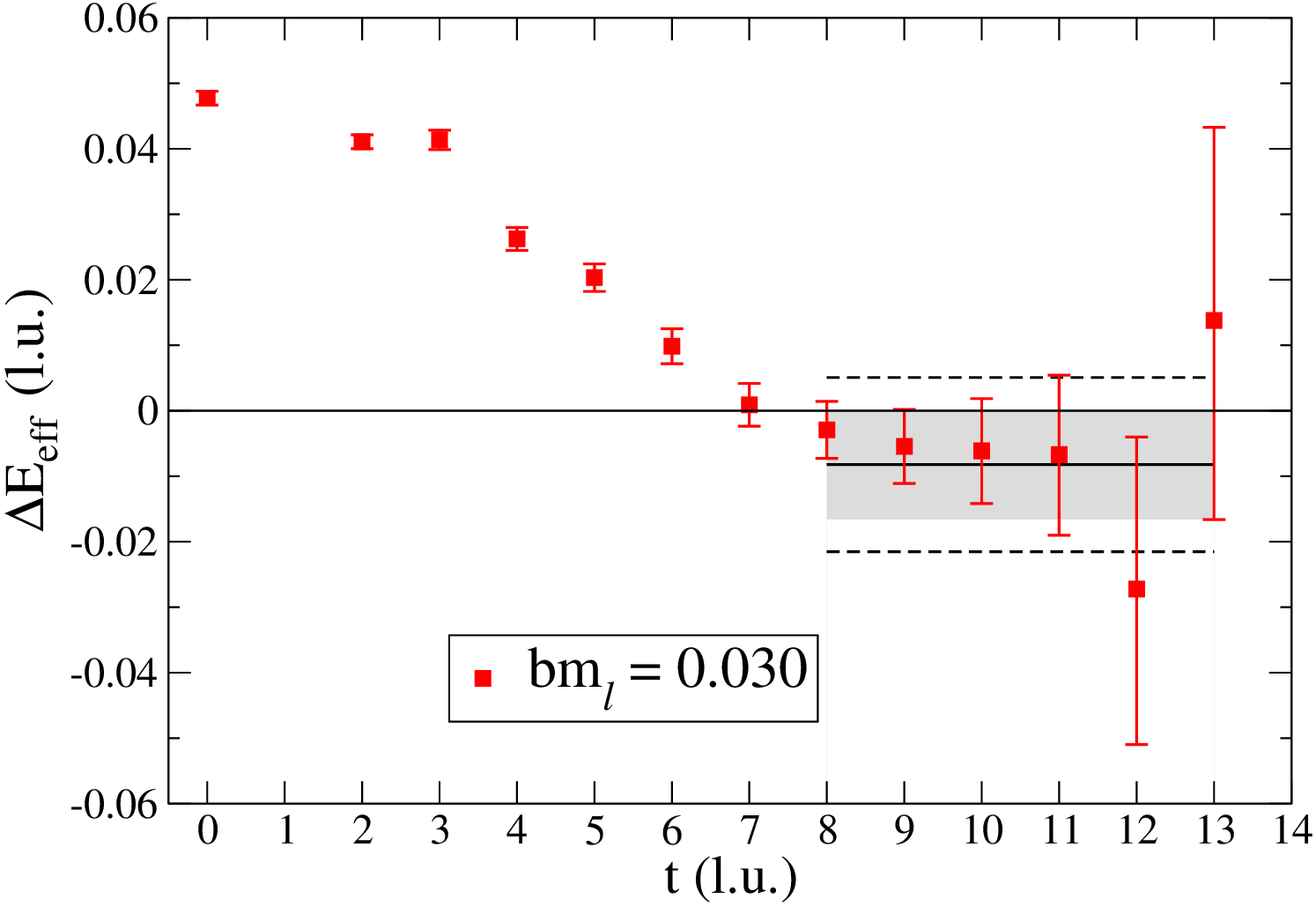} 
\caption{Effective mass plots for  $n\Lambda$ in the
  $\siii$-channel
at pion masses of $m_\pi\sim 350~{\rm MeV}$ (left panel),
$m_\pi\sim 490~{\rm MeV}$ (center panel) and $m_\pi\sim
590~{\rm MeV}$ (right panel).
The straight line and shaded region correspond to the extracted energy shift
and associated uncertainty. The dashed lines correspond to the statistical
and systematic errors added linearly. }
\label{fig:effmassNL3s1}
\end{figure}

\begin{figure}[!ht]
\vskip0.2in
\includegraphics[width = .49\textwidth,angle=-0] {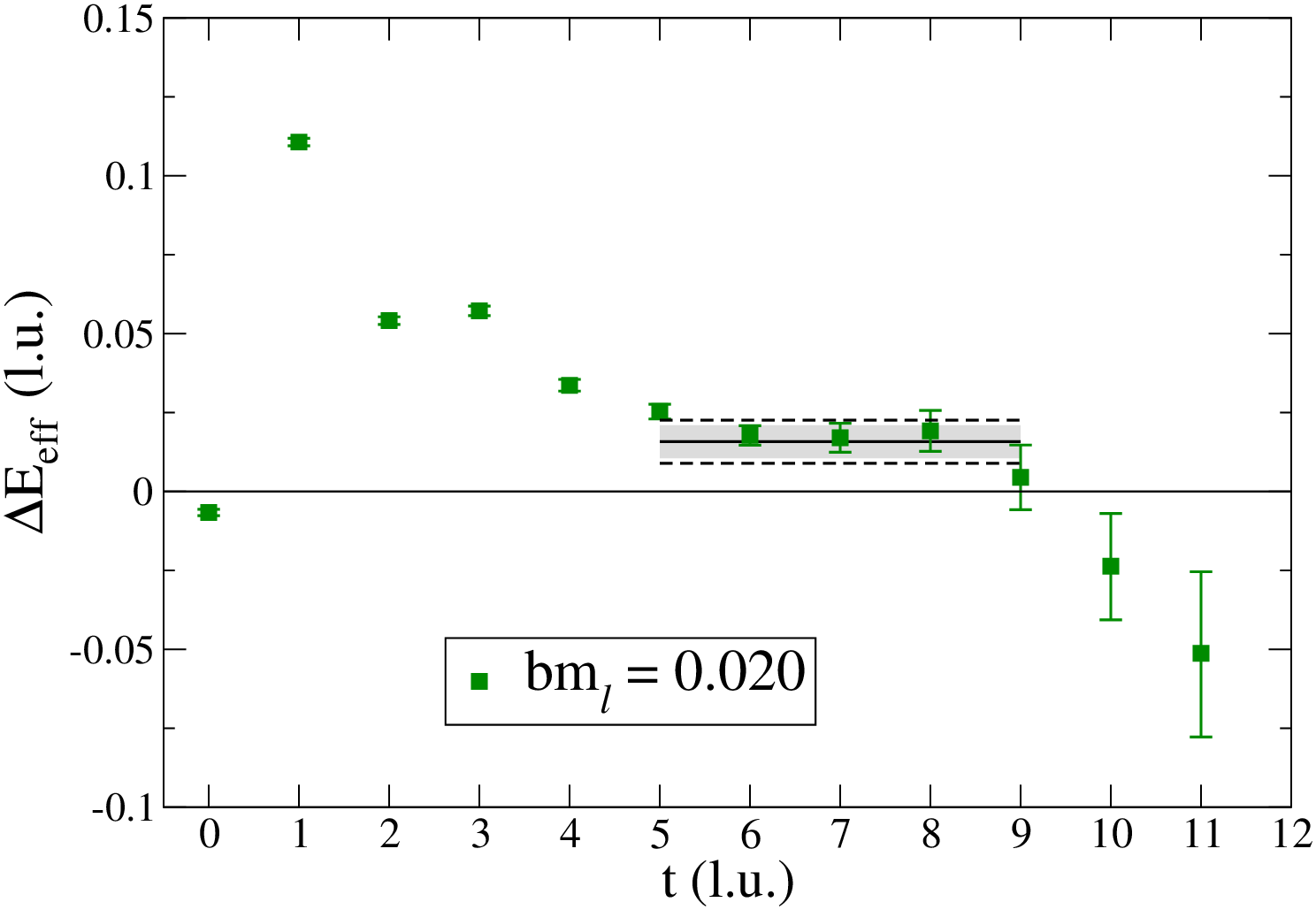}\hfill
\includegraphics[width = .49\textwidth,angle=-0] {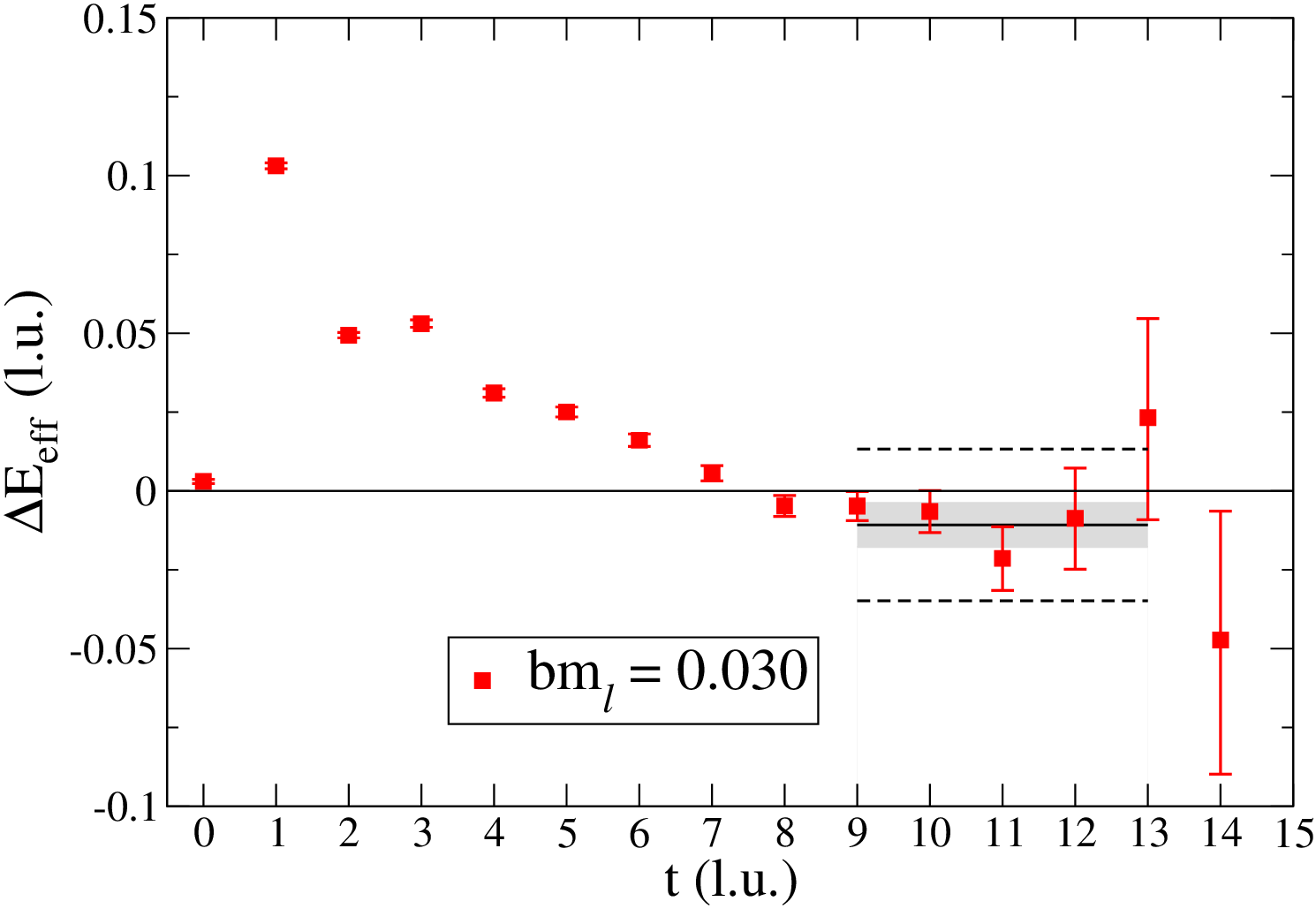}
\caption{Effective mass plots for  $n\Sigma^-$ in the
  $\si$-channel
at pion masses of $m_\pi\sim 490~{\rm MeV}$ (left panel) and $m_\pi\sim
590~{\rm MeV}$ (right panel).
The straight line and shaded region correspond to the extracted energy shift
and associated uncertainty. The dashed lines correspond to the statistical
and systematic errors added linearly.
}
\label{fig:effmassNS1s0} 
\end{figure}

\begin{figure}[!ht]
\includegraphics[width = .49\textwidth,angle=-0] {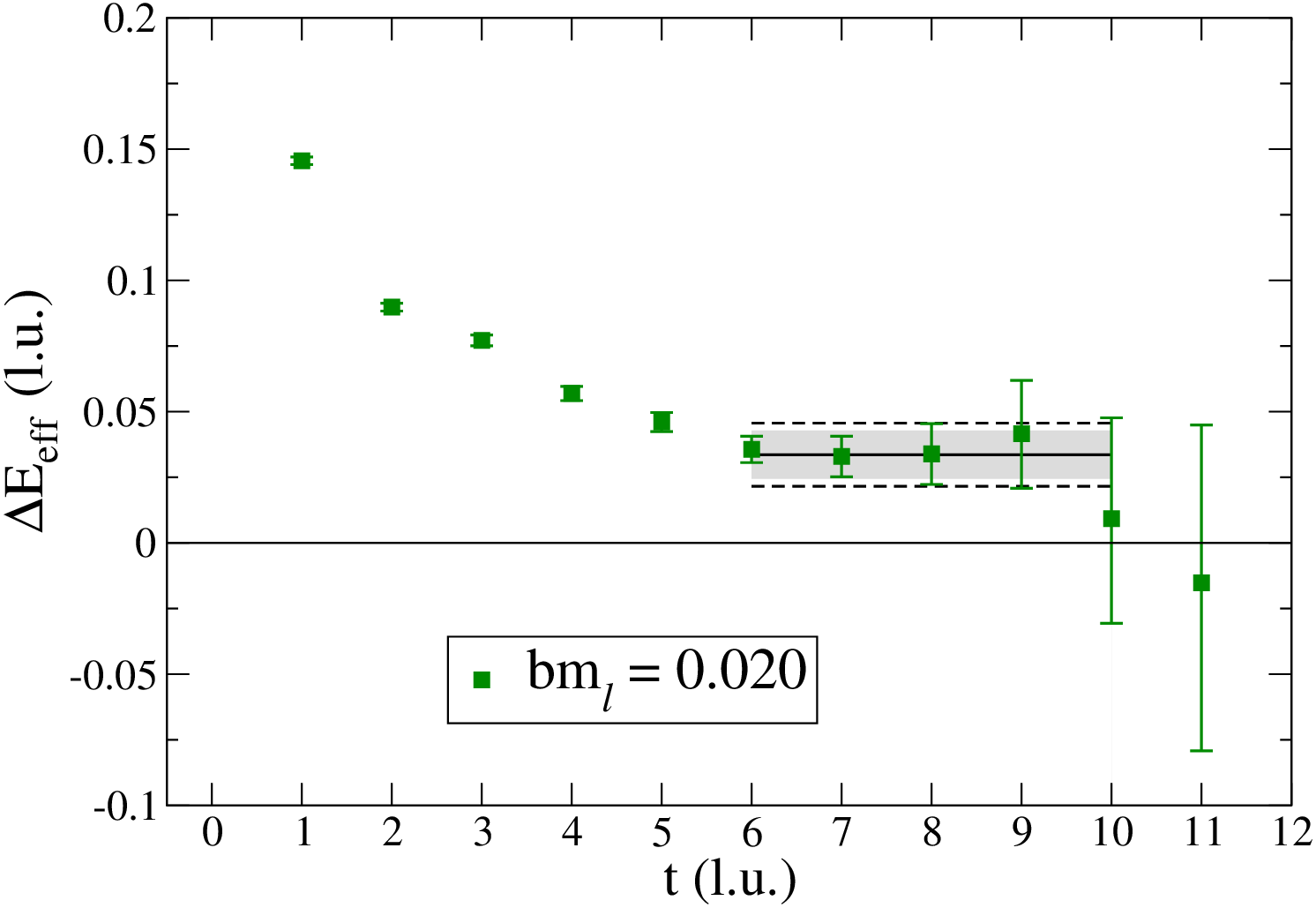}\hfill
\includegraphics[width = .49\textwidth,angle=-0] {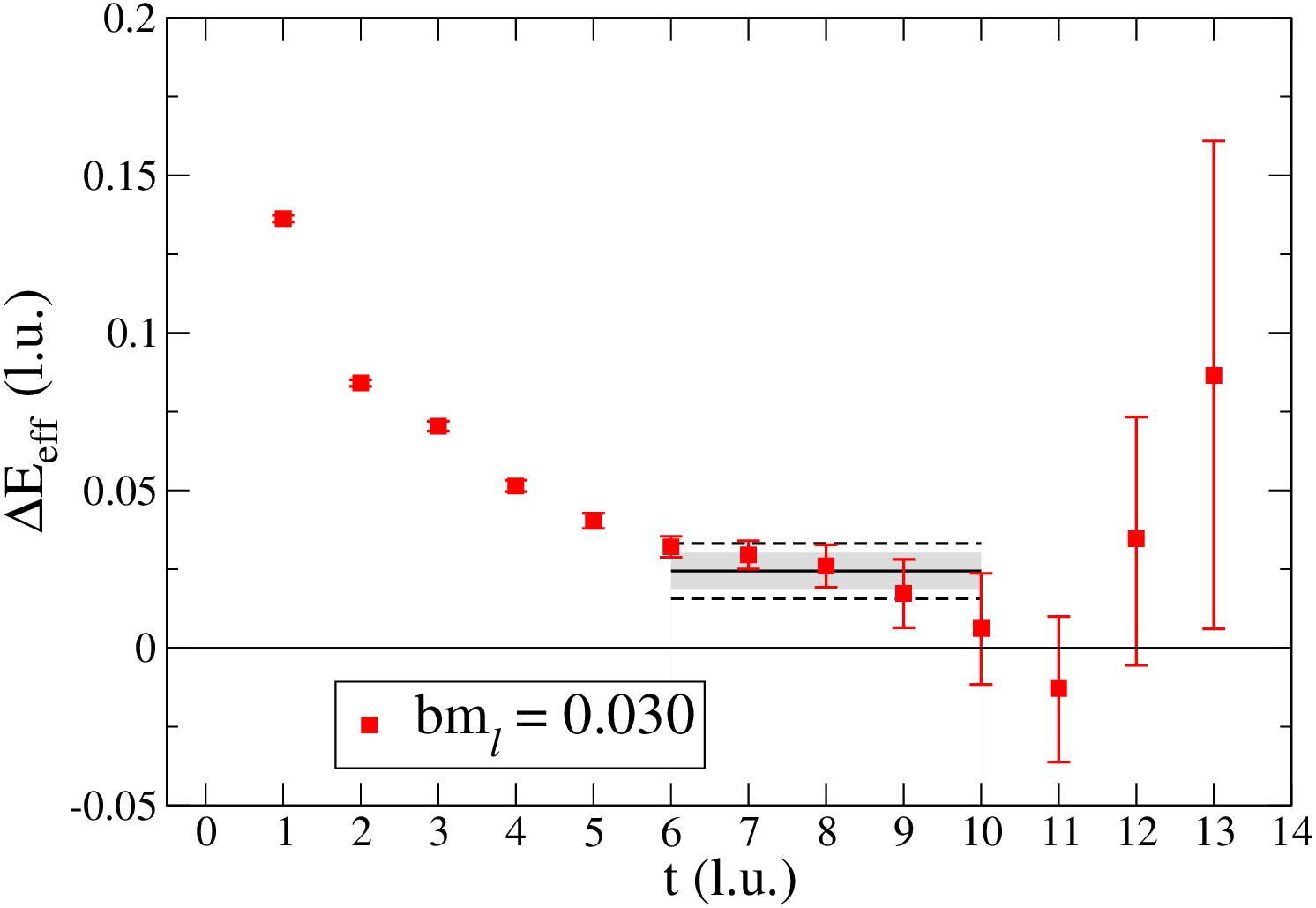}
\caption{Effective mass plots for  $n\Sigma^-$ in the
  $\siii$-channel
at pion masses of $m_\pi\sim 490~{\rm MeV}$ (left panel) and $m_\pi\sim
590~{\rm MeV}$ (right panel).
The straight line and shaded region correspond to the extracted energy shift
and associated uncertainty. The dashed lines correspond to the statistical
and systematic errors added linearly.
}
\label{fig:effmassNS3s1}
\end{figure}

\noindent The effective mass plots of the ratio of correlation functions with
identifiable plateaus obtained at $m_\pi\sim 350~{\rm MeV}$,
$m_\pi\sim 490~{\rm MeV}$ and $m_\pi\sim 590~{\rm MeV}$ are shown in
figs.~\ref{fig:effmassNL1s0}, \ref{fig:effmassNL3s1},
\ref{fig:effmassNS1s0} and \ref{fig:effmassNS3s1}.
%
%
%
%
%
%
%
Single and double exponential forms were fit to the correlation
functions by $\chi^2$-minimization, from which the $YN$ energy shifts
were determined.  The central values and statistical uncertainties were determined
by the jackknife procedure over the ensemble of configurations, and
are shown in Table~\ref{table:1}.
%
%
\begin{table}
\begin{center}
\begin{tabular}{|c|c|c|c|c|c|c|}
\hline
Channel 
& $m_\pi$ (MeV)  
& Range   
& $\Delta E$ (MeV)    
& $|{\bf k}|$ (MeV) 
&  $\delta$ (degrees)     
& $-(k\cot\delta)^{-1}$ (fm)\\
\hline
\hline
$n\Lambda$  
&\ $592\pm 1\pm 10$ \  
&\  8-12\   
&\  $-9\pm 8\pm 20\ $\ 
& -- \ 
&\ -- \  
&\ $0.8\pm 1.4\pm 0.4$ \ 
\\
\cline{2-7}
$\si$  
& $493 \pm 1 \pm 8$ 
& 6-9 
& \ $29.8\pm 5.4\pm 2.5$ \ 
& $197\pm 24\pm 4$ 
&\  $-32.3\pm 8.1\pm 2.8$ \ 
& \ $0.63\pm 0.12 \pm 0.014$\ 
\\
\cline{2-7} 
&  $354\pm 1\pm 6$ 
&  5-9  
& $56.8\pm 6.0\pm 5.5$   
&\  $255\pm 22\pm 13$\ 
& $-53.4\pm 8.5 \pm 10.1$  
& $1.04\pm 0.24\pm 0.15$ 
\\
\hline
\hline
$n\Lambda$ 
& \ $592\pm 1\pm 10$  
& 8-13  
& $-13\pm 13\pm 8$ 
& --  
& --  
& $3\pm 14\pm 2$ 
\\
\cline{2-7}
$\siii$ 
& $493\pm 1\pm 8$ 
& 7-11 
& $-4\pm 13 \pm 14$   
& --  
& --  
& $(-\infty,\infty)$ 
\\
\cline{2-7}
& $354\pm 1\pm 6$ 
& 5-10 
& $23\pm 17\pm 4$   
& $168\pm 62\pm 14$  
& $-23\pm 18\pm 4$  
& $0.50\pm
0.26\pm 0.06$ \\
\hline
\hline
$n\Sigma^-$
& \ $592\pm 1\pm 10$  
& 9-13 
& $-17\pm 11\pm 27$   
& --  
& -- 
& $(-\infty,\infty)$ 
\\
\cline{2-7}
$\si$ 
&  $493 \pm 1 \pm 8$ 
& 5-9 
& $24.9\pm 7.8\pm 3.0$ 
& $ 179\pm 28\pm 11$ 
& $-27.2\pm 9.0\pm 3.8$  
& $0.57\pm 0.13\pm 0.05$\\
\hline
\hline
$n\Sigma^-$ 
& \ $592\pm 1\pm 10$  
& 6-10  
& $38.5\pm 8.8\pm 5.0$ 
& $ 226\pm 26\pm 15$ 
& $-44.3\pm 9.8\pm 5.4$  
& $0.85\pm 0.20\pm 0.10$ 
\\
\cline{2-7}
$\siii$ 
& $493 \pm 1 \pm 8$ 
& 6-10 
& $53\pm 14 \pm 5$ 
& $261\pm 35\pm 13$ 
& $-58\pm 15\pm 5$  
& $1.19\pm
0.51\pm 0.15$ \\
\hline
\end{tabular}
\end{center}
\caption{\label{table:1}
Summary of results from the $G_{YN}$
correlation functions which exhibit a clear plateau in the effective energy
plot. The first error is statistical and the second error is systematic.}
\end{table}

The plateaus in the correlator ratios $G_{YN}(t)$ persist for only a
small number of time-slices.  At small $t$ there is the usual
contamination from excited states whereas at larger $t$ the
signal-to-noise ratio degrades exponentially with $t$ according to
$e^{-(M_N+M_Y-3m_\pi)t}$\cite{lepage_error}.  The Dirichlet boundary
at $t=22$ introduces a systematic uncertainty due to backward
propagating states.  However, in practice, the statistical noise
becomes a limiting factor at far earlier time slices and the boundary
at $t=22$ is not an issue for this calculation.  We obtained a
non-zero energy shift larger than the statistical error in ten of the
$G_{YN}$ correlation functions, as shown in Table~\ref{table:1}.  The
phase shifts $\delta$ (and $-1/k\cot\delta(k)$) were determined
through the L\"uscher formula and their errors by the jackknife
procedure (we do not give a value of $\delta$ for a negative energy shift).
The quantities in Table~\ref{table:1} that are in physical units were
obtained with a lattice spacing of $b=0.125$ fm set by MILC, which is
consistent with the determination from \cite{nplqcd-pipi} ($b=0.1274
\pm 0.0007 \pm 0.0003$ fm) based on the chiral expansion of the pion decay
constant.  We have not shown the results for channels in which there
is no clear plateau in the effective mass plot. This is the case for
all $m_\pi\sim 290~{\rm MeV}$ ($m_l=0.007$) correlation functions
where the rapid decrease of the signal-to-noise ratio caused by the
small pion mass eliminated all plateaus. The systematic errors shown
in Table~\ref{table:1} are determined by varying the fitting range,
and by comparing the results of fitting one and two exponentials to
the ratio of correlation functions.

It is not clear that we have been able to identify the ground states
in all of the correlation functions, e.g. $n\Sigma^-$ in the
$\si$-channel at $m_\pi\sim 490~{\rm MeV}$, and $n\Lambda$ in the
$\si$-channel at $m_\pi\sim 490~{\rm MeV}$, as the statistics are not
sufficient to determine whether the large-time behavior we observe is due
to noise or due to the presence of any states with lower energy than
those shown in Table~\ref{table:1}. Indeed, it would be very exciting
if there were states with lower energy, as they would likely be bound
states (based on naturalness arguments and the exact L\"uscher
relation).  This uncertainty in no way undermines our results;
regardless of the nature of the states shown in Table~\ref{table:1},
they are clearly YN states present in the continuum.

In addition to the fitting systematics given in Table~\ref{table:1},
there are other systematic uncertainties in our calculations that we
have not shown, as they are all expected to be small in comparison.
The discretization errors due to the finite lattice spacing arising
from the sea-action are of order $\mathcal{O}(\alpha_s b^2)$
($\alpha_s$ is the strong coupling constant) and those in the valence
sector of order $\mathcal{O}(b^2)$ due to the near-perfect chiral
symmetry.  The agreement between continuum chiral perturbation theory
and other results based on the same discretization scheme
\cite{nplqcd-pipi,Beane:2006mx,nplqcd-go,nplqcd-isobreak} strongly
suggest that the discretization errors in the hyperon interactions
are, at most, of the order of a few percent. This is much smaller than
the statistical errors quoted in Table \ref{table:1}.  A similar point
can be made regarding the use of different fermion actions in the
valence and sea sectors. Further, the relation between energy levels
and phase shifts, eq.~(\ref{eq:energies}), is valid only up to
corrections that are exponentially small in the volume.  The
corrections to the L\"uscher formula can be computed in chiral
perturbation theory, as shown in the $\pi\pi$ case in
\cite{bedaque_finitepipi} and for two nucleons in
\cite{bedaque_finiteNN}.  These effects are particularly small in the
$N\Lambda$ system, as the long-range part of the interaction is
dominated by two-pion exchange and one-kaon exchange, and not one-pion
exchange.
\begin{figure}[!ht]
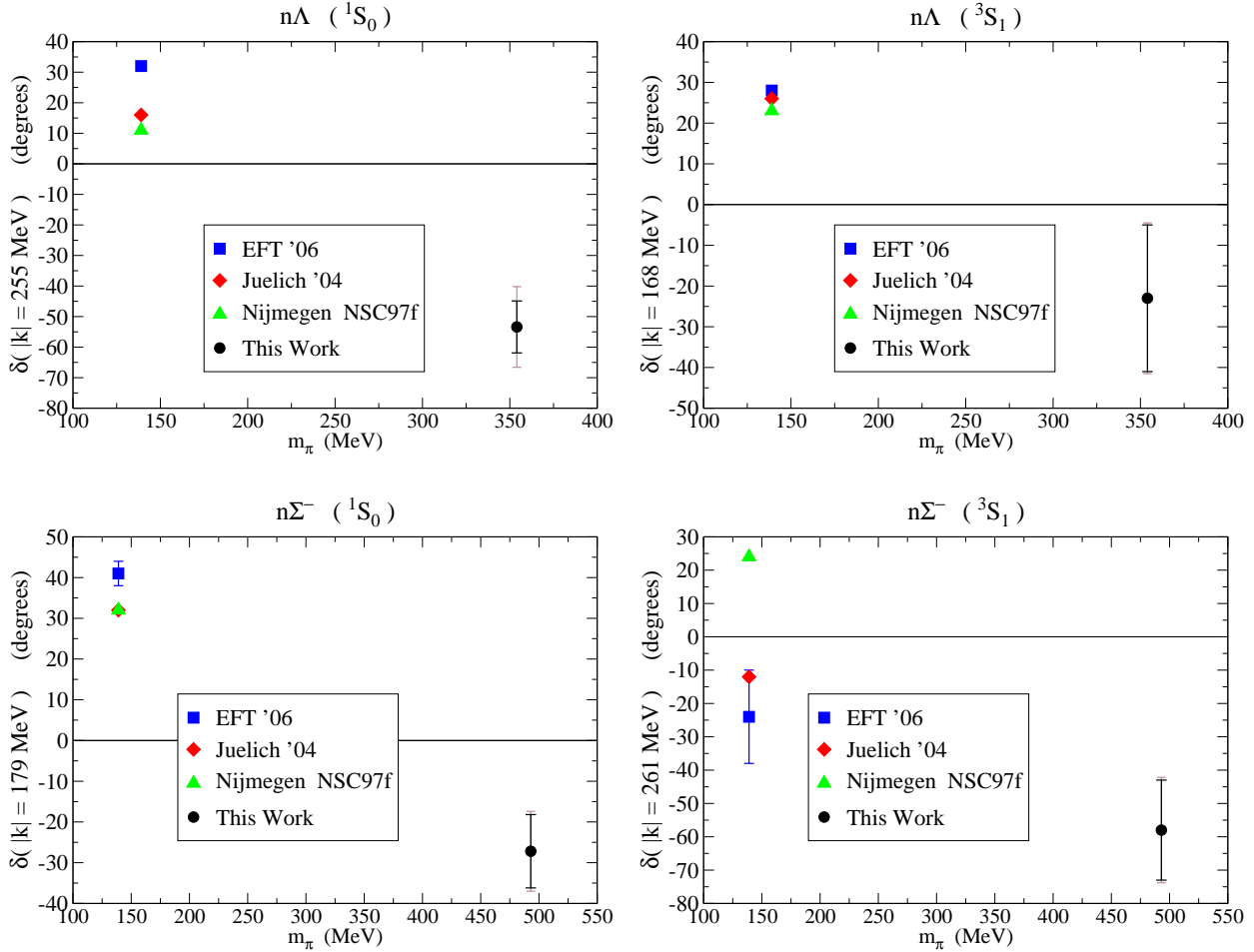

\vskip.1in
\includegraphics[width = .49\textwidth,angle=0] {NLamSING_354.eps} \hfill
\includegraphics[width = .49\textwidth,angle=0] {NLamTRIP_354.eps}
\vskip.2in
\includegraphics[width = .49\textwidth,angle=0] {NSigSING_493.eps} \hfill
\includegraphics[width = .49\textwidth,angle=0] {NSigTRIP_493.eps}
\caption{Comparison of the lowest-pion-mass lattice results in each channel with
a recently developed YN EFT~\cite{PHM06} (squares), and several potential
models: Nijmegen~\cite{nij99} (triangles) and J\"ulich~\cite{HM05}
(diamonds). The dark error bars on the lattice data are statistical and the light error bars
are statistical and systematic errors added in quadrature.}
\label{fig:phaseshcompare}
\end{figure}
%

\subsection{Discussion}

\noindent We have presented results of the first fully-dynamical
lattice QCD calculation of YN interactions.  The scattering amplitudes
for s-wave $n\Lambda $ and $n\Sigma^-$, in both the $\si$-channel and
the $\siii-\diii$ coupled-channels, have been determined at one value
of momentum for pion masses of $\sim 350~{\rm MeV}$, $\sim 490~{\rm
MeV}$ and $590~{\rm MeV}$.  Unfortunately, the lightest pion mass at
which we have been able to extract a signal is at the upper limits of
the regime of applicability of the effective field theories that have
been constructed, thus precluding a chiral extrapolation.  However, this work does provide new rigorous
theoretical constraints on effective field theory, and potential model
constructions of YN interactions.  In fig.~\ref{fig:phaseshcompare} we
compare the lattice values of the phase shifts to recent EFT
results~\cite{PHM06} (squares), and to several potential models:
Nijmegen~\cite{nij99} (triangles) and J\"ulich~\cite{HM05}. At
face value these results appear quite discrepant, however one should keep in
mind that extrapolation to the physical pion mass will seriously alter
individual contributions to the YN interaction.

While the measurements of the momenta and phase shifts are
unambiguous, their physical interpretation is not entirely resolved.
Each of the phase shifts at the lowest pion masses are negative.
Assuming that the observed state is the ground state in the lattice
volume, this implies that the interactions are all repulsive.  The
$n\Sigma^-$ interaction in the $\siii-\diii$ coupled channels is
strongly repulsive at $m_\pi\sim 490~{\rm MeV}$, while the interaction
in the $\si$-channel is only mildly repulsive.  The opposite is found
to be true for the $n\Lambda$ systems at $m_\pi\sim 350~{\rm MeV}$,
where the interaction in the $\si$-channel is found to be strongly
repulsive, while that in the $\siii-\diii$ coupled channels is mildly
repulsive.  However, there may be channels for which there exist
states of lower, negative energies, some of which may correspond to
bound states in the continuum limit.  If such states are present, then
we would conclude that the interaction is attractive, and that the
positive-shifted energy state we have identified corresponds to the
first continuum level.  Current statistics are sufficiently poor that
nothing definitive can be said about the existence of such
states. Therefore, we are continuing to accumulate statistics and
experiment with signal optimization in order to resolve this issue.

It is clear that a precise lattice QCD calculation of YN scattering
will have dramatic impact upon the field of hypernuclear physics, and
may have an equal impact on our ability to determine the evolution of
neutron stars, simply due to the present absence of precise
experimental data.  We have performed the first of such calculations,
albeit at unphysically large pion masses. The present work was
limited entirely by the lack of computational resources.  We hope that
this limitation recedes in the future, and that lattice QCD can be
developed as a reliable tool to calculate the interactions between
baryons in experimentally inaccessible or difficult areas of strong
interactions.

\subsection{Acknowledgments}

\noindent We thank R.~Edwards for help with the QDP++/Chroma
programming environment~\cite{Edwards:2004sx} with which the
calculations discussed here were performed. The computations for this
work were performed at Jefferson Lab, Fermilab, Lawrence Livermore
National Laboratory, National Center for Supercomputing
Applications, Centro Nacional de Supercomputaci\'on (Barcelona, Spain)
and ASTRON-BlueGene/L at the Reken Centrum of Groningen University.
We are indebted to the MILC and the LHP collaborations
for use of their configurations and propagators, respectively.  The
work of MJS was supported in part by the U.S.~Dept.~of Energy under
Grant No.~DE-FG03-97ER4014. The work of KO was supported in part by
the U.S.~Dept.~of Energy contract No.~DE-AC05-06OR23177 (JSA) and
contract No.~DE-AC05-84150 (SURA).  The work of PFB was supported in
part by the U.S.~Dept.~of Energy grant No.~ER-40762-365. The work of
SRB was supported in part by the National Science Foundation under
grant No.~PHY-0400231.  Part of this work was performed under the
auspices of the US DOE by the University of California, Lawrence
Livermore National Laboratory under Contract No. W-7405-Eng-48. AP is
supported by the Ministerio de Educaci\'on y Ciencia (Spain) under
contract No.~FIS2005-03142 and by the Generalitat de Catalunya under
contract No.~2005SGR-00343.

\end{document}